# Subwavelength grating waveguide filter based on cladding modulation with phase-change material grating

S. Hadi Badri[1], Saeid Gholami Farkoush[2,*]

[1] Department of Electrical Engineering, Sarab Branch, Islamic Azad University, Sarab, Iran

[2] Department of Electrical Engineering, Yeungnam University, 280 Daehak-Ro, Gyeongsan, Gyeong-sangbuk-do 38541, Korea

* saeid_gholami@yu.ac.kr

## Abstract

Subwavelength engineering and utilizing phase-change materials with large contrast in their optical properties have become powerful design tools for integrated silicon photonics. Reversible phase-transition of phase-change materials such as $Ge_2Sb_2Te_5$ (GST) provides a new degree of freedom and opens up the possibility of adding new functionalities to the designed devices. We present an optical filter based on a silicon subwavelength grating (SWG) waveguide evanescently coupled to phase-change material loading segments arranged periodically around the SWG core. The effect of the GST loading segments' geometry and their distance from the SWG core on the filter's central wavelength and bandwidth are studied with three-dimensional finite-difference time-domain simulations. The employment of GST in the structure adds a switching functionality with an extinction ratio of 28.8 dB. We also examine the possibility of using the proposed structure as a reconfigurable filter by controlling the partial crystallization of the GST offering a blueshift of more than 4 nm.

## 1. Introduction

Subwavelength grating (SWG) structures provide us with new approaches to design photonic integrated circuits (PIC) components due to their ability to tailor the modal confinement, effective refractive index, dispersion, and birefringence of the structure [1, 2]. In SWG waveguides, the core is composed of periodically interlacing silicon pillars with low-index material at the subwavelength scale while it is surrounded by a cladding material with a lower refractive index [3]. SWG structure can operate in the radiation, Bragg reflection, and subwavelength regimes. The Bragg wavelength of the structure compared to the free-space wavelength of the light determines the operation regime of the structure. The Bragg wavelength is defined as $\lambda_B=2n_{eff}\Lambda$ where $\Lambda$ and $n_{eff}$ are the grating period and the effective refractive index of the structure, respectively. When the wavelength of the light is considerably longer than the Bragg wavelength, $\lambda>>\lambda_B$, the structure is in the radiation regime where the light partially radiates out of the waveguide. Optical couplers operate in this regime [4, 5]. In the Bragg reflection regime, the propagation of light is prohibited, due to reflection, for wavelengths near the Bragg wavelength. Various devices have been proposed that operate in the Bragg reflection regime [6-8]. In the subwavelength regime, the wavelength of the light is considerably smaller than the Bragg wavelength, therefore, the diffraction and reflection effects are suppressed and the structure can be regarded as an effective medium governed by effective medium theory [9, 10]. Various components have been proposed that operate in the subwavelength regime [11, 12]. Spectral filtering is one of the important components of the integrated photonics and various structures have been utilized to design filters such as Bragg grating structures [13-16] and metal-insulator-metal structure with metallic grating [17, 18]. These devices do not have any other functionality. Recently, phase-change materials (PCMs) have been introduced to photonics enabling new functionalities and reconfigurability in the photonic devices. PCMs have two phases with a large contrast in their optical properties. The reversible phase-transition of PCMs has been utilized to design various devices such as optical switches [19, 20], modulators [21, 22], memories [23, 24], and various metasurfaces [25-29]. Other materials with tunable optical properties such as graphene [30] and Dirac semimetal $Cd_3As_2$ [31] have been investigated to design tunable optical devices. However, the devices designed by these materials are volatile with a drawback of static power consumption which is required to keep the state of the device. PCMs such as GST are non-volatile reducing the static power consumption of the reconfigurable devices considerably.

In this paper, we employ a silicon SWG waveguide loaded with phase-change material loading segments to add filtering and switching functionalities to the designed structure. GST is a non-volatile phase-change material that has two reversible phases with large optical/electrical contrast between them [32]. The amorphous and crystalline phases of GST can be induced by optical [33, 34], electrical [35, 36], or thermal [37, 38] excitations. By tuning the geometrical parameters of the GST loading segments and their distance from the SWG core, we can control the coupling strength and the rejection band of the structure. Phase-transition of the GST segments from amorphous to crystalline state enables us to switch off the transmitted light. By partial crystallization of GST, we can achieve intermediate states with unique optical properties estimated by effective medium theory. These intermediate states may be used to design a reconfigurable filter enabling us to shift its central wavelength. Here, we focus on the amplitude modulation of the designed structure, however, the phase modulation based on PCMs have also been studied [39, 40].

## 2. SWG waveguide with GST loading segments

The proposed structure is shown in Fig. 1. The silicon SWG waveguide, with a grating pitch of $\Lambda$=275 nm, operates in the subwavelength regime. The width of the silicon SWG structure is $w$=500 nm while its duty cycle is $D$=0.7, therefore, the length of silicon pillars is $L= D\times\Lambda$ =192.5 nm. A taper with a length of $L_{taper}$=10×Λ=2.75 μm is employed to convert the mode of the strip waveguide to the Floquet–Bloch mode of the SWG waveguide. The periodic placement of cubic GST loading segments along the silicon SWG is utilized to design a spectral filter. The grating period of the loading segments is $\Lambda_L$=470 nm with a duty cycle of $D_L$=0.7. The side length of the GST cubes is $L_L= D_L\times\Lambda_L$ =329 nm. The length of the Bragg grating is $L_{BG}$=17.86 μm. The gap between the GST loading segments and the silicon SWG waveguide is $g_L$=100 nm. Moreover, the longitudinal shift of the Bragg grating at the two sides of the SWG waveguide is $s_L$=0. The substrate is silica while the upper cladding is air.

The central wavelength of the filter is $\lambda_c=2n_{eff}\Lambda_L$ where $n_{eff}$ is the effective refractive index of the structure. The bandwidth and the maximum reflection of the filter are determined by [41]

$$\Delta\lambda = \frac{\lambda_c^2}{\pi n_g}\sqrt{\kappa^2 + \frac{\pi^2}{L_{BG}{}^2}} \tag{1}$$

$$R_{\max} = \tanh^2(\kappa L_{BG}) \tag{2}$$

where $\kappa$ is the coupling coefficient representing the coupling rate between forward and backward propagating modes of the structure. And $n_g$ is the group index. These equations indicate that the characteristics of the filter depend on the index modulation and the length of the grating ($L_{BG}$). Moreover, $\kappa$ and $n_g$ depend largely on the grating geometry [42]. In our design, the silicon SWG waveguide operates in the subwavelength regime and it can be regarded as an effective medium while the GST loading segments operate in the Bragg reflection regime. Therefore, the geometry of the GST loading segments plays the main role in determining the bandwidth, central wavelength, and maximum reflection of the structure. In the next section, the effect of phase-transition of GST loading segments as well as their geometrical parameters on the transmission spectrum of the proposed structure is investigated.

The proposed GST grating can be fabricated by the following procedure. The GST is deposited on $SiO_2$ by RF co-sputtering from stoichiometric GeTe and $Sb_2Te_3$ targets. To protect the GST from oxidation and to confine heat in the structure during the switching process a protective layer should be added on top of the GST. $SiO_2$ deposited by the atomic layer deposition method or $Si_2N_4$ deposited by DC sputtering from an Si target in a reactive Ar:$N_2$ atmosphere can be used as a protective layer. Then electron-beam lithography (EBL) and reactive ion etching (RIE) are used to get the desired pattern [20, 43].

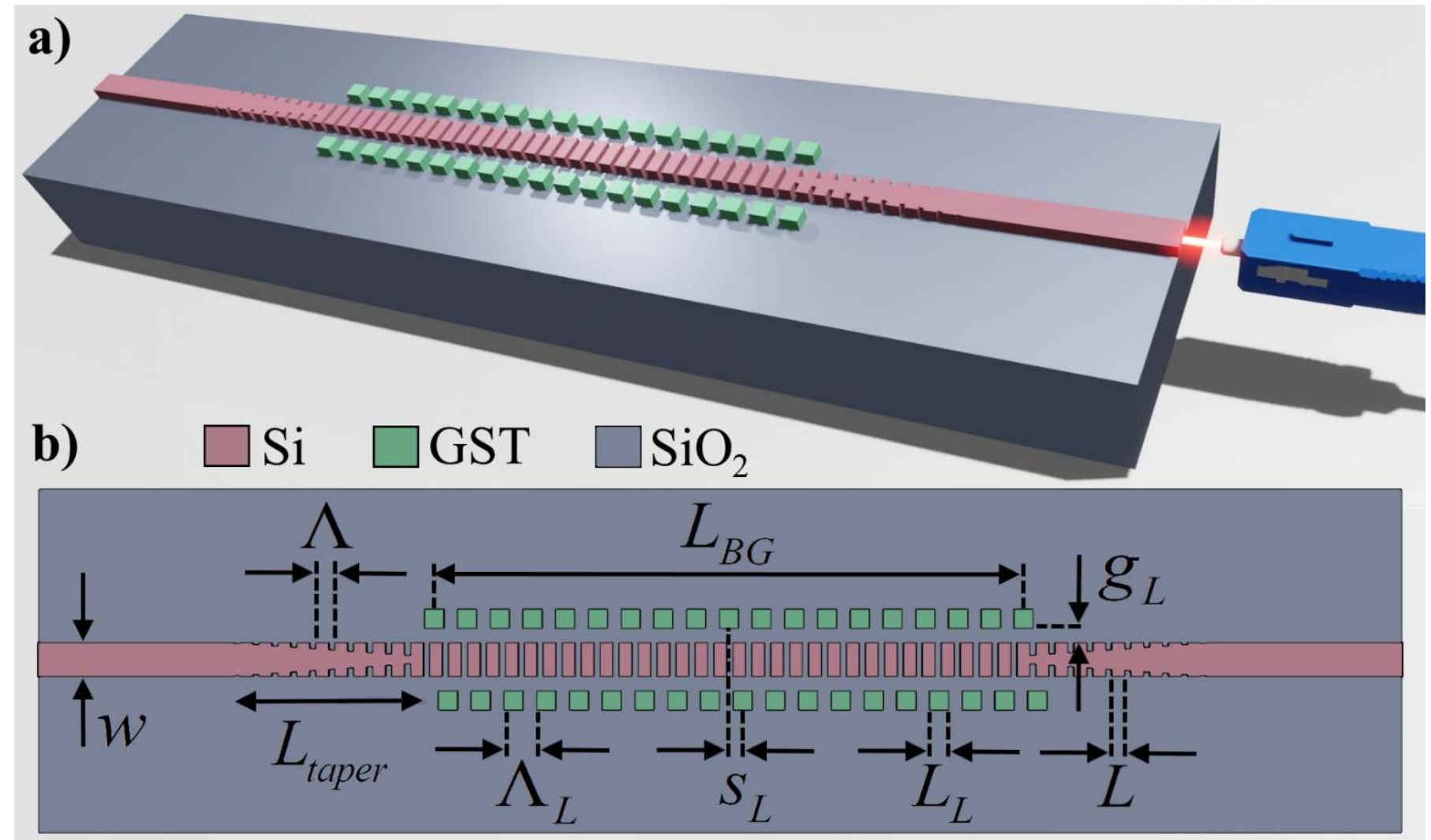

Fig. 1. a) The structure of the optical filter. b) Top-view of the structure. The upper cladding is air.

## 3. Results and discussion

We utilized three-dimensional (3D) finite-difference time-domain (FDTD) simulations to evaluate the performance of the designed structure. The optical properties of Si and $SiO_2$ defined by Palik were used in our simulations [44]. For the amorphous and crystalline GST, the wavelength dependent refractive index and extinction coefficient provided in [25] were employed. For each structure, we perform the simulations twice. First, the optical properties of the amorphous GST are used in the calculations and for the second run of simulations, the crystalline GST's properties are used. In the simulations, we suppose that the phase-transition has been induced by optical, electrical, or thermal excitation and the structure is in steady state condition and the room temperature. We terminated the computational domain by the perfectly matched layer (PML) in all directions. A non-uniform meshing with a minimum meshing step of 20 nm was used in all directions. In this section, we examine the effect of phase-transition of GST and geometrical parameters of the structure on the filtering and switching functionalities of the proposed structure. In all the simulations, the transverse electric (TE) mode is injected into the waveguide. Throughout the discussions of this section, the geometrical parameters of the silicon SWG waveguide are fixed to $\Lambda$=275 nm, $w$=500 nm, $L$=192.5, and $L_{taper}$=2.75 µm.

### 3.1 Effect of the duty cycle

We first examine the effect of the loading segments' duty cycle ($D_L$) on the transmission spectrum of the device. As the duty cycle increases the length of GST segments increases leading to the increase in the effective refractive index of the structure ($n_{eff}$). This, in turn, increases $\lambda_B$. As can be seen in Fig. 2(a), when GST segments are in the amorphous state, the increase in the duty cycle results in the shift of the central wavelength of the filter to longer wavelengths. For $D_L$=0.5, the resonance is centered near 1526.85 nm with the minimum transmission of 0.30 while the full-width half-maximum (FWHM) is about 28.1 nm. In calculating the FWHM, we first determine the attenuation contrast of the filter by calculating the difference between transmission level in the passband and the minimum transmission level in the rejection band. Adding the half of the filter's attenuation contrast to the minimum transmission in the rejection band, we find the transmission level for calculating FWHM. The quality factor of the filter is defined as Q= $\lambda_c$/FWHM. In this case, the quality factor of the filter is about 54. Increasing the duty cycle of loading segments to $D_L$=0.6 redshifts the central wavelength to 1539.71 nm with a minimum transmission of 0.26. In this case, the FWHM is about 29.6 nm while the quality factor is about 52. For $D_L$=0.7, the transmission minimum occurs at the wavelength of 1555.33 nm with a transmission of 0.29 while the FWHM is 27.6 nm and the quality factor of about 56. As can be seen in Fig. 2(b), transition of the GST segments to the crystalline state decreases the transmission of the structure considerably. For $D_L$=0.5, the transmission is lower than 0.42 in the wavelength range of 1400-1700 nm. In this case, there is a trough in the transmission spectrum at the wavelength of 1518.56 nm with a transmission of 0.1 indicating that the proposed structure may be used as a reconfigurable filter which will be discussed in subsection 3.6. For $D_L$=0.6, the transmission of the structure is lower than 0.15. For the larger duty cycle of $D_L$=0.7, the transmission is lower than $T_{off}$=0.001, therefore, the structure effectively acts as an optical switch. Considering that the maximum transmission in the passband of the structure, while the GST segments are in the amorphous state, is about $T_{on}$=0.77, we can calculate the extinction ratio of the switch as $ER = 10\log(T_{on}/T_{off})$=28.8 dB. Since the structure with $D_L$=0.7 provides a better switching functionality, we chose $D_L$=0.7 for our structure.

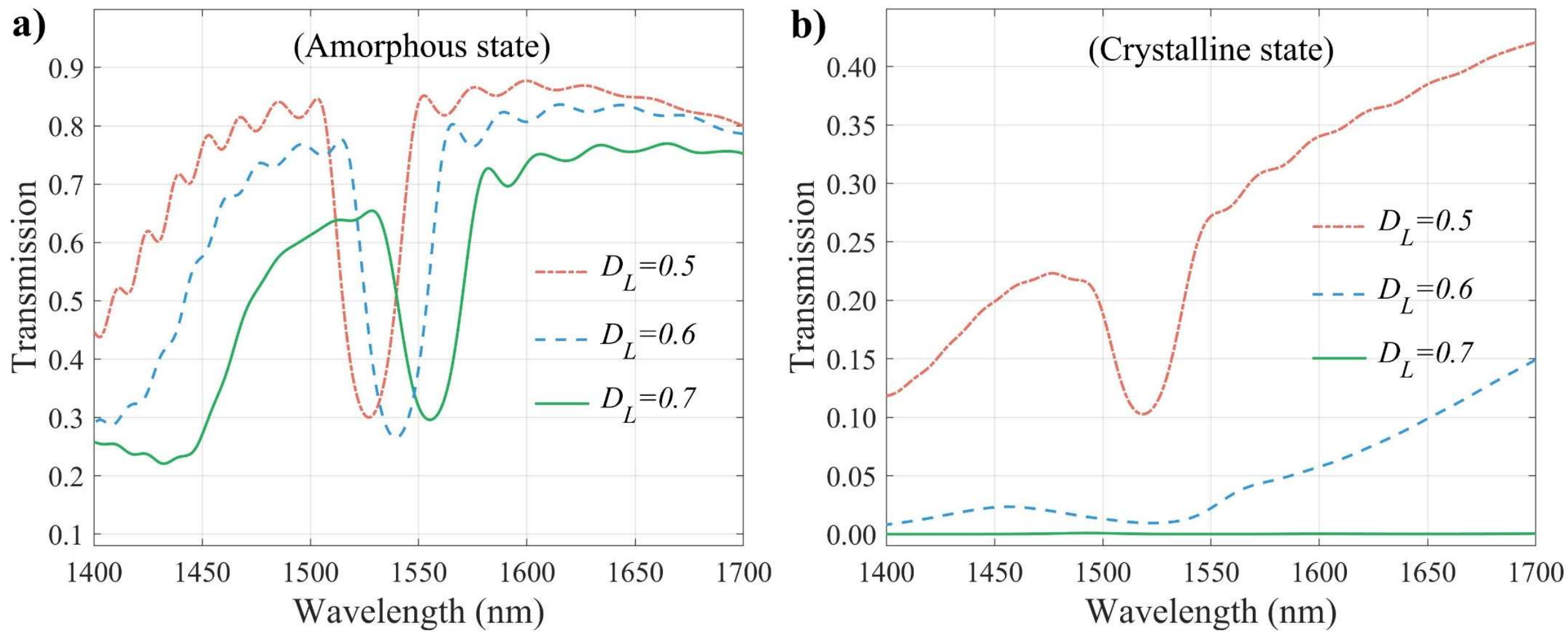

Fig. 2. The effect of duty cycle of the GST segments ($D_L$) on the transmission spectra of the designed structure when the GST segments are in the a) amorphous and b) crystalline states. Other geometrical parameters of the loading segments are fixed to $\Lambda_L$=470 nm, $s_L$= 0, $L_{BG}$=17.86 µm, and $g_L$=100 nm.

### 3.2 Effect of the Bragg grating's length

In this subsection, we examine the effect of the Bragg grating's length ($L_{BG}$) on the performance of the structure. The central wavelength depends on the grating pitch and the effective refractive index of the structure ($\lambda_B$=2$n_{eff}\Lambda_L$), therefore, the structures with different lengths of the grating have a central wavelength of about 1555 nm. However, as indicated by Eqs. 1 and 2, the bandwidth and the maximum reflection of the filter depend on the Bragg grating's length. As shown in Fig. 3(a), when the GST segments are in the amorphous state, for a shorter grating with $L_{BG}$=12.22 µm, the transmission is 0.45 at the trough while the FWHM is about 30.6 nm with the quality factor of about 51. As the length of the grating increases to $L_{BG}$=17.86 µm, the maximum reflection increases, and consequently, the transmission at the trough decreases to 0.29. Moreover, according to Eq. 1, increasing the length of the Bragg grating results in decease of the bandwidth. For $L_{BG}$=17.86 µm, the FWHM is about 27.6 nm while the quality factor is about 56. Further increase of the grating's length to $L_{BG}$=23.50 µm yields better performance with a transmission of 0.17 at the trough of the transmission spectrum with the FWHM of about 24.2 nm and the quality factor of about 64. As shown in Fig. 3(b), when the GST segments are in the crystalline state, the structure hinders the transmission of light significantly. As the length of the Bragg grating increases the transmission of the structure decreases. For $L_{BG}$=12.22 µm, the maximum transmission is lower than 0.008. Increasing the length of the Bragg grating to $L_{BG}$=17.86 µm decreases the transmission of the structure to lower than 0.001. For $L_{BG}$=23.50 µm, the transmission of the structure is negligible. A longer Bragg grating provides superior performance regarding the filtering and switching functionalities of the proposed structure, however, to reduce both memory requirements and simulation time, we chose $L_{BG}$=17.86 µm.

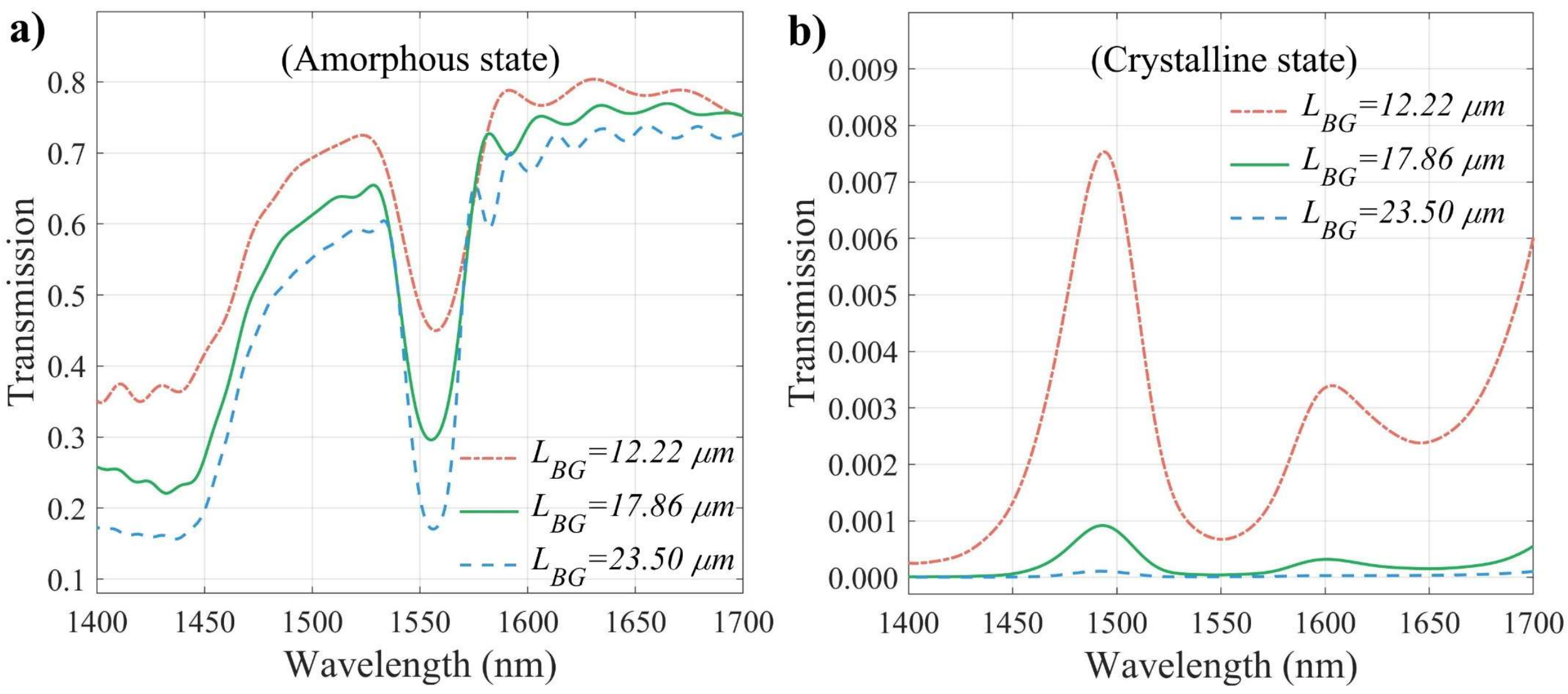

Fig. 3. The effect of length of the Bragg grating composed of GST segments on the transmission spectra of the designed structure when the GST segments are in the a) amorphous and b) crystalline states. Other geometrical parameters of the loading segments are fixed to $\Lambda_L$=470 nm, $s_L$= 0, $L_L$=329 nm, and $g_L$=100 nm.

### 3.3 Effect of the longitudinal shift of GST segments

The grating strength can also be controlled by a longitudinal shift of the Bragg grating at the two sides of the SWG waveguide [45]. The coupling coefficient can be described as $\kappa = \kappa_0 \cos(\pi \times s_L / \Lambda_L)$ where $\kappa_0$ is the coupling coefficient for the structure without misalignment of loading segments [46]. When the misalignment between the GST loading segments is $s_L$=0, the light is strongly reflected back at the Bragg wavelength ($\lambda_B$). However, as the longitudinal shift increases to half the grating period ($s_L$=$\Lambda_L$/2), the light transmits through the structure without reflection. As shown in Fig. 4(a), when the GST loading segments are in the amorphous state, the transmission trough occurs at the wavelength of 1555.33 nm with a transmission of 0.29 for $s_L$=0. However, increasing the misalignment to $s_L$=100 nm leads to the decrease of reflection at the Bragg wavelength of 1554.83 nm and increases the transmission at this wavelength to about 0.4. Increasing the longitudinal shift to $s_L$= 200 nm, further decreases the reflection, and consequently, the transmission increase to about 0.64 at the wavelength of 1552.29 nm. As shown in Fig. 4(b), the phase transition of the GST segments to the crystalline state results in the dramatic decrease of the transmitted light. For the longitudinal shift of $s_L$=0, the transmission is lower than 0.001. Increasing the longitudinal shift of the GST loading segments improves the switching functionality of the structure. However, considering the filtering functionality of the structure, we chose $s_L$=0.

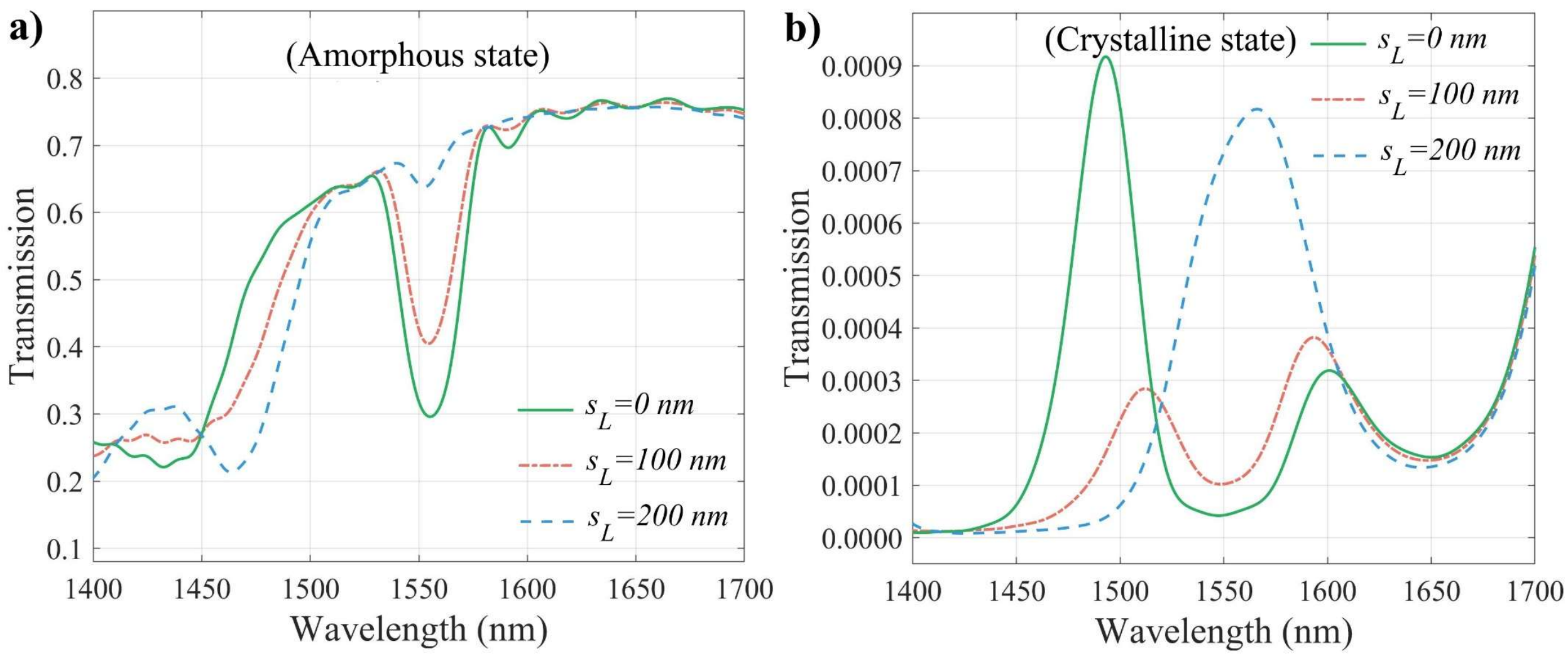

Fig. 4. The effect of longitudinal shift of the GST segments ($s_L$) on the transmission spectra of the designed structure when the GST segments are in the a) amorphous and b) crystalline states. Other geometrical parameters of the loading segments are fixed to $\Lambda_L$=470 nm, $L_L$= 329 nm, $L_{BG}$=17.86 µm, and $g_L$=100 nm.

### 3.4 Effect of the gap between the SWG waveguide and GST loading segments

The gap between the silicon SWG waveguide and the loading segments ($g_L$) determines the grating strength [45]. As the gap increases the grating strength decreases considerably since a smaller portion of the evanescent wave interacts with the loading segments. Moreover, the effective refractive index of the structure decreases as the gap increases leading to the blueshift of the Bragg wavelength. The effect of $g_L$ on the transmission of the filtering functionality of the structure, while the GST loading segments are in the amorphous state, is shown in Fig. 5(a). For $g_L$=100 nm, the resonance is centered near the wavelength of 1555.33 nm with a transmission of about 0.29. Increasing the gap to $g_L$=200 nm blueshifts the Bragg wavelength to 1516.13 nm while the transmission is about 0.68 at this wavelength. The grating strength of the loading segments decreases considerably as the gap increases to $g_L$=300 nm. In this case, there is no bandstop in the transmission spectrum of the structure. Phase transition of the GST segments to crystalline state decrease the transmission of the structure as shown in Fig. 5(b). For $g_L$=100 nm, the transmission is lower than 0.001. Increasing the gap to $g_L$=200 nm, increases the transmission of the structure to 0.045. However, for the $g_L$=300 nm, the switching functionality of the structure decreases considerably and the transmission increases to a maximum of 0.25. Considering the filtering and switching ability of the structure at smaller gaps, we chose $g_L$=100 nm.

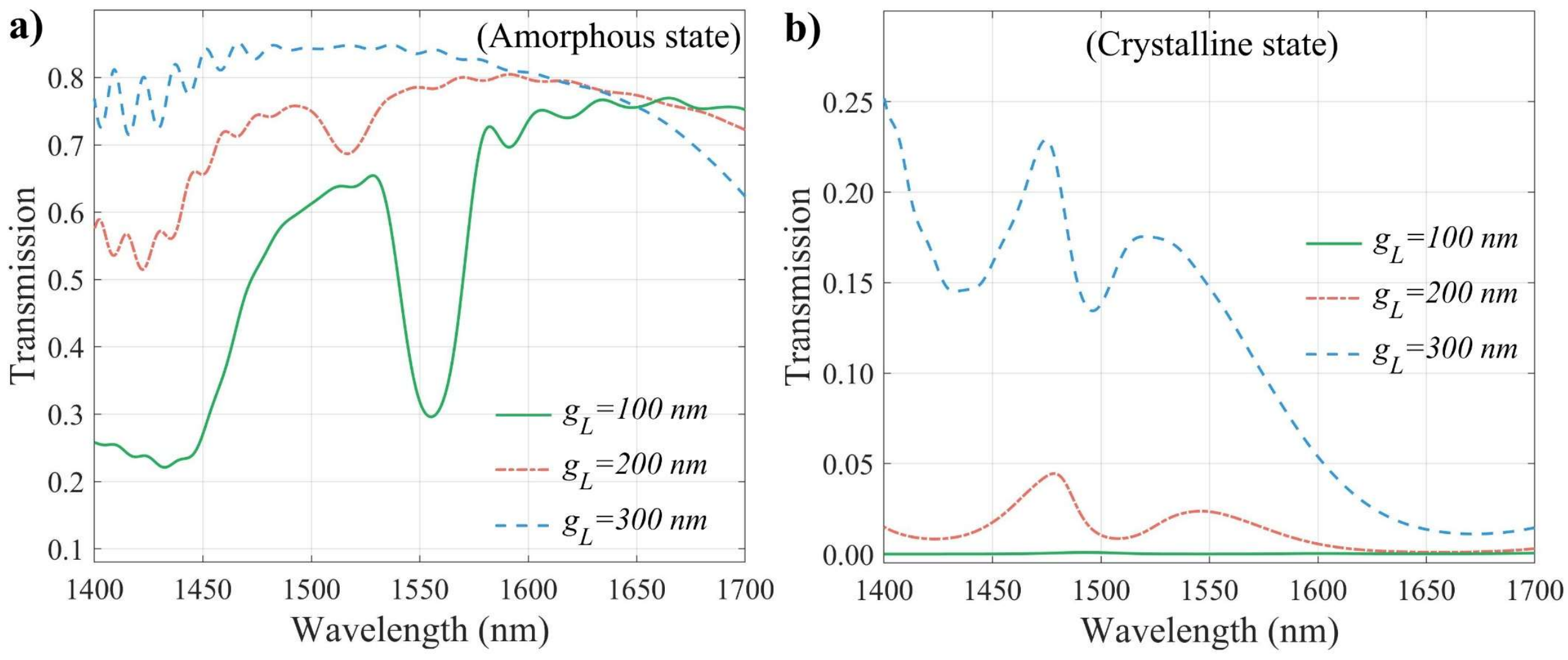

Fig. 5. The effect of gap between the SWG waveguide and GST segments ($g_L$) on the transmission spectra of the designed structure when the GST segments are in the a) amorphous and b) crystalline states. Other geometrical parameters of the loading segments are fixed to $\Lambda_L$=470 nm, $L_L$= 329 nm, $L_{BG}$=17.86 µm, and $s_L$=0.

### 3.5 Effect of grating pitch

While the GST segments are in the amorphous state, the central wavelength of the filter is directly proportionate to the grating pitch of GST loading segments as displayed in Fig. 6(a). For $\Lambda_L$=450 nm, the Bragg wavelength of the structure occurs at the wavelength of 1514.68 nm with a transmission of 0.35. In this case, the FWHM is about 24.6 and the quality factor is about 61. By increasing the grating pitch of GST loading segments to $\Lambda_L$=470 nm, the Bragg wavelength redshifts to 1555.33 nm with a transmission of about 0.29 while the FWHM is about 27.6 nm and the quality factor is about 56. For $\Lambda_L$=490 nm, the central wavelength of the bandstop shifts to a longer wavelength of 1598.23 nm with a transmission of about 0.24. The FWHM is 30.6 nm and the quality factor is about 52. As shown in Fig. 6(b), the phase transition of GST segments to the crystalline state switches off the transmitted light. For $\Lambda_L$=450 nm, the transmission is lower than 0.006 while for $\Lambda_L$=470 nm, the transmission is lower than 0.001. For $\Lambda_L$=490 nm, the transmission in the wavelength range of 1400-1700 nm range is lower than 0.0006. We chose $\Lambda_L$=470 nm since the central wavelength of the filter is closer to the center of the C-band of optical communications.

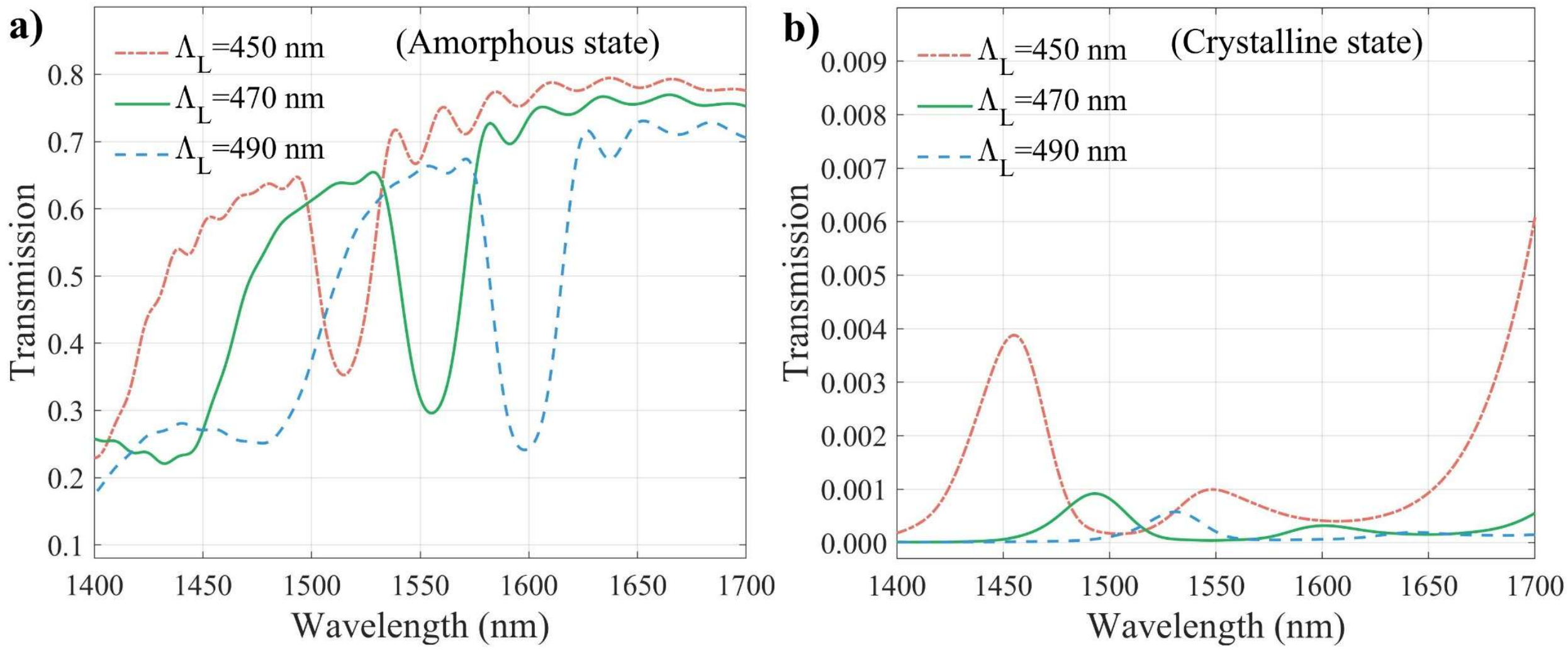

Fig. 6. The effect of grating pitch of GST segments ($\Lambda_L$) on the transmission spectra of the designed structure when the GST segments are in the a) amorphous and b) crystalline states. Other geometrical parameters of the loading segments are fixed to $L_L$= 0.7×$\Lambda_L$, $L_{BG}$=17.86 µm, $s_L$=0, and $g_L$=100 nm.

### 3.6 The proposed structure as a reconfigurable filter

In this subsection, we examine the effect of partial crystallization of GST segments on the performance of the filter. By heating the amorphous GST, its temperature gradually approaches the crystallization point while numerous small crystalline puddles are formed. Finally, these puddles join and the GST becomes crystalline. The intermediate phases composed of amorphous and crystalline puddles can be realized and the optical properties of these states are estimated by the effective medium theory (EMT). Various composite structures have been utilized to design novel photonic components based on the EMT [47-49]. The effective permittivity ($\varepsilon_{eff}$) of the intermediate state is estimated by the Lorentz-Lorenz relationship [50]

$$\frac{\varepsilon_{eff}(\lambda)-1}{\varepsilon_{eff}(\lambda)+2}=C\frac{\varepsilon_c(\lambda)-1}{\varepsilon_c(\lambda)+2}+(1-C)\frac{\varepsilon_a(\lambda)-1}{\varepsilon_a(\lambda)+2} \quad (3)$$

where $C$ is the crystallization fraction. $\varepsilon_a(\lambda)$ and $\varepsilon_c(\lambda)$ are the wavelength-dependent permittivities of the amorphous and crystalline GST, respectively. As shown in Fig. 7, the proposed structure can be used as a reconfigurable filter. Here, the geometrical parameters of the GST loading segments are $\Lambda_L$=470 nm, $L_L$= 235 nm, $L_{BG}$=17.86 µm, $s_L$=0, and $g_L$=100 nm. In the amorphous state, the central wavelength of the filter is located at 1526.85 nm with the transmission of 0.30 while the stopband width of the filter is about 29 nm. When 25% of the GST is crystallized, the Bragg wavelength blueshifts to 1524.89 nm with a transmission of about 0.24. In this case, the bandwidth of the stopband is about 30 nm. Increasing the crystallization of the GST to 50%, the Bragg wavelength blueshifts to 1522.45 nm with a transmission of 0.19 while the its FWHM is about 29.3 nm with a quality factor of about 52. It should be noted that at higher crystallization fractions, the extinction coefficient of the GST segments increases leading to lower transmission levels at the passbands of the filter. Consequently, the attenuation contrast of the filter decreases considerably. Therefore, we limit the crystallization fraction to 50% which in turn limits the achievable wavelength shift of the filter. Using a PCM with higher refractive index change upon phase-transition and lower extinction coefficient can increase the turnability of the designed filter. However, this would decrease the extinction ratio of the structure as an optical switch. Considering this trade-off between the dual functionalities of the designed structure, GST was chosen as the PCM.

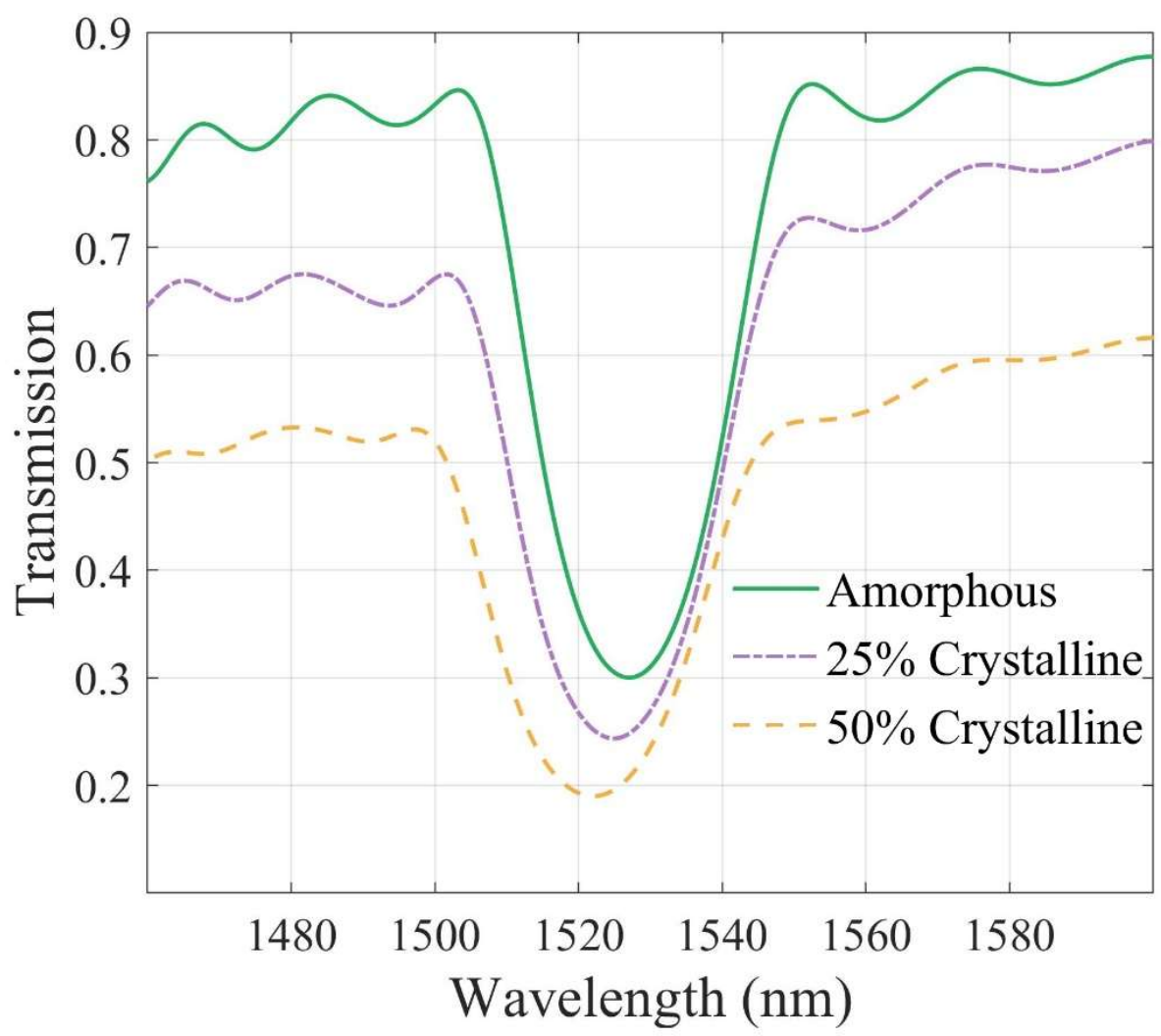

Fig. 7. The effect of partial crystallization of GST segments on the transmission spectra of the designed structure. The geometrical parameters of the loading segments are $\Lambda_L$=470 nm, $L_L$= 235 nm, $L_{BG}$=17.86 µm, $s_L$=0, and $g_L$=100 nm.

For a structure with $\Lambda_L$=470 nm, $L_L$= 329 nm, $L_{BG}$=17.86 µm, $s_L$=0, and $g_L$=100 nm, the distribution of the electric field intensity at the operating wavelengths in the amorphous and crystalline states of GST is displayed in Fig. 8. While the GST is in the amorphous state, the injected light passes through the filter with low return loss in the passbands of the filter as shown in Fig. 8(a). For instance, at the wavelength of 1600 nm, 1.7% of the injected light to the filter is reflected while 73% is transmitted. We place monitors near the PML boundaries to estimate the scattered light. At 1600 nm, about 8% of the injected light is scattered and, therefore, about 17.3% is absorbed by the lossy GST. At the central wavelength of the stopband (1555 nm) 29.6% of the injected light is transmitted while 45.6% is reflected. About 8% of the injected light is scattered, hence, about 16.8% is absorbed by the GST segments. In the crystalline state of GST, 21.4% is reflected while the transmitted power is about 0% at the wavelength of 1555 nm. Moreover, about 8% of the injected light is scattered, therefore, 70% is absorbed by the lossy GST segments.

We estimated the effect of imperfections in the fabrication of GST segments by introducing ±20 nm random deviations in the size and position of the GST segments. In the amorphous state, such deviations in the structure may shift the central wavelength of the structure by up to 3 nm while the transmission level in the passbands may decrease by up to 0.1. In the crystalline state, the fabrication imperfections have limited effect on the performance of the switch and the transmission remains below 0.001.

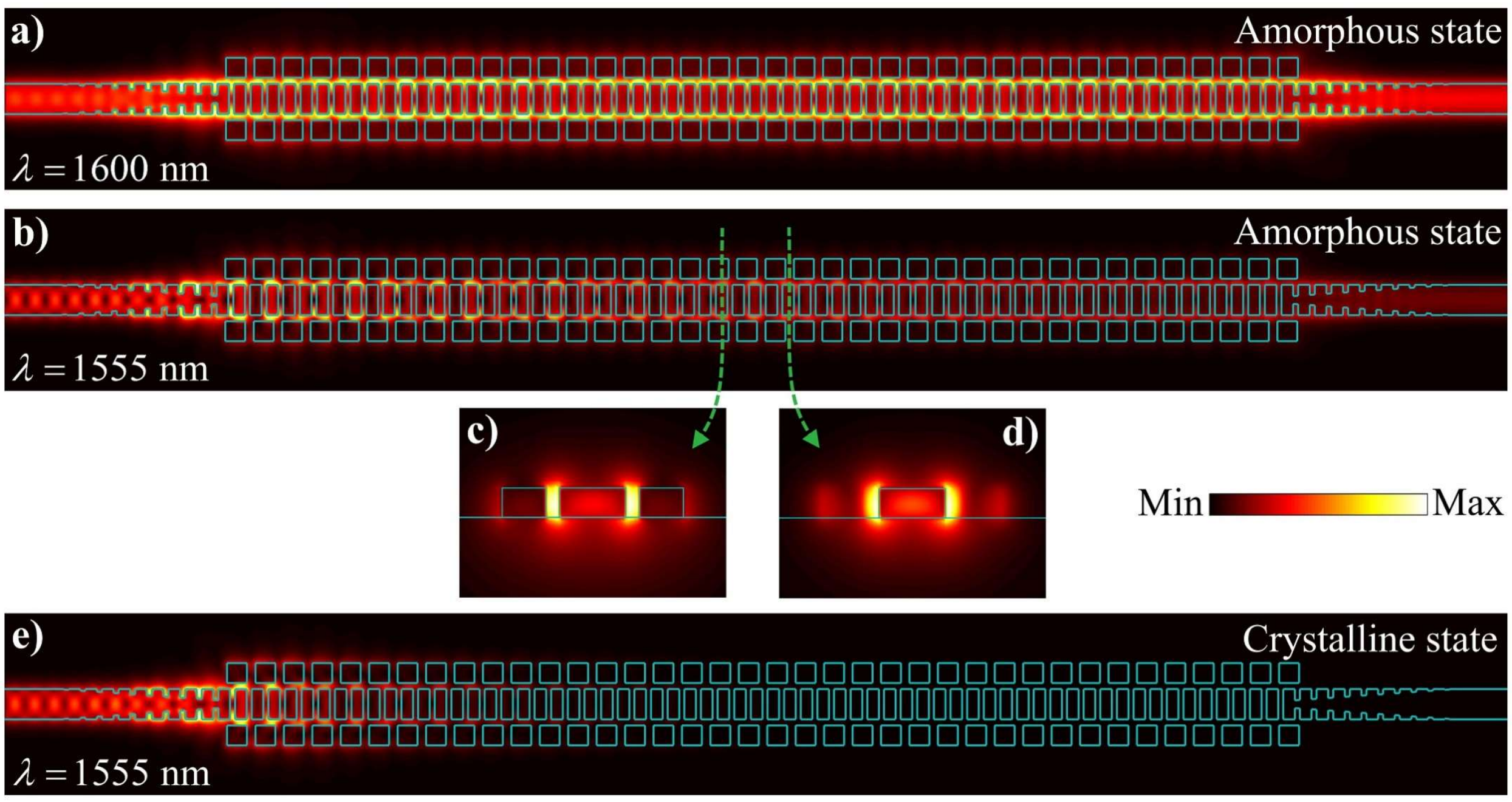

Fig. 8. The electric field intensity of the structure in the amorphous state at the wavelengths of a) 1600 nm and b) 1555 nm. The electric field intensity in two interfaces of the structure are shown in c) and d). e) The electric field intensity in the amorphous state of the GST at the wavelength of 1555 nm.

## 4. Conclusion

We present an optical filter based on a silicon SWG waveguide where its effective refractive index is modulated by loading segments placed around the waveguide. The GST has been chosen as the loading segments operating as a Bragg reflector while the silicon SWG structure operates in the subwavelength regime treated as an effective medium. The depth and width of transmission dip of a notch filter is controlled by the geometry of GST loading segments and their distance from the waveguide's core. The 3D FDTD simulations were employed to evaluate the spectral characteristics and study their dependences on the geometrical parameters of the loading segments. Simulations indicate that the phase-transition of the GST can introduce switching to the designed structure. When the GST is in the amorphous state, the structure operates as a filter while in the crystalline state, the transmission is lower than 0.001 in the bandwidth of 1400-1700 nm. Moreover, the central wavelength of the filter can be reconfigured by controlling the partial crystallization of the GST.

**Disclosures.** The authors declare no conflicts of interest.